\documentclass{vgtc}                          

\graphicspath{{figures/}{pictures/}{images/}{./}} 

\usepackage{times}                     

\usepackage{tabu}                      
\usepackage{booktabs}                  
\usepackage{lipsum}                    
\usepackage{mwe}                       

\usepackage{mathptmx}                  
\usepackage{amsmath}
\usepackage{enumitem}
\usepackage{url}

\onlineid{7155}

\vgtccategory{Research}

\vgtcinsertpkg

\title{Spatial Affordance-aware Interactable Subspace Allocation for \\ Mixed Reality Telepresence}

\author{Dooyoung Kim\thanks{e-mail: dooyoung.kim@kaist.ac.kr}\\ %
        \scriptsize KAIST%
\and Seonji Kim\thanks{e-mail: seonji.kim@kaist.ac.kr}\\ %
     \scriptsize KAIST%
\and Selin Choi\thanks{e-mail: selin.choi@kaist.ac.kr}\\ %
     \scriptsize KAIST %
\and Woontack Woo\thanks{e-mail: wwoo@kaist.ac.kr}\\ %
     \parbox{1.4in}{\scriptsize \centering KAIST}}

\teaser{
  \centering
  \includegraphics[width=\linewidth]{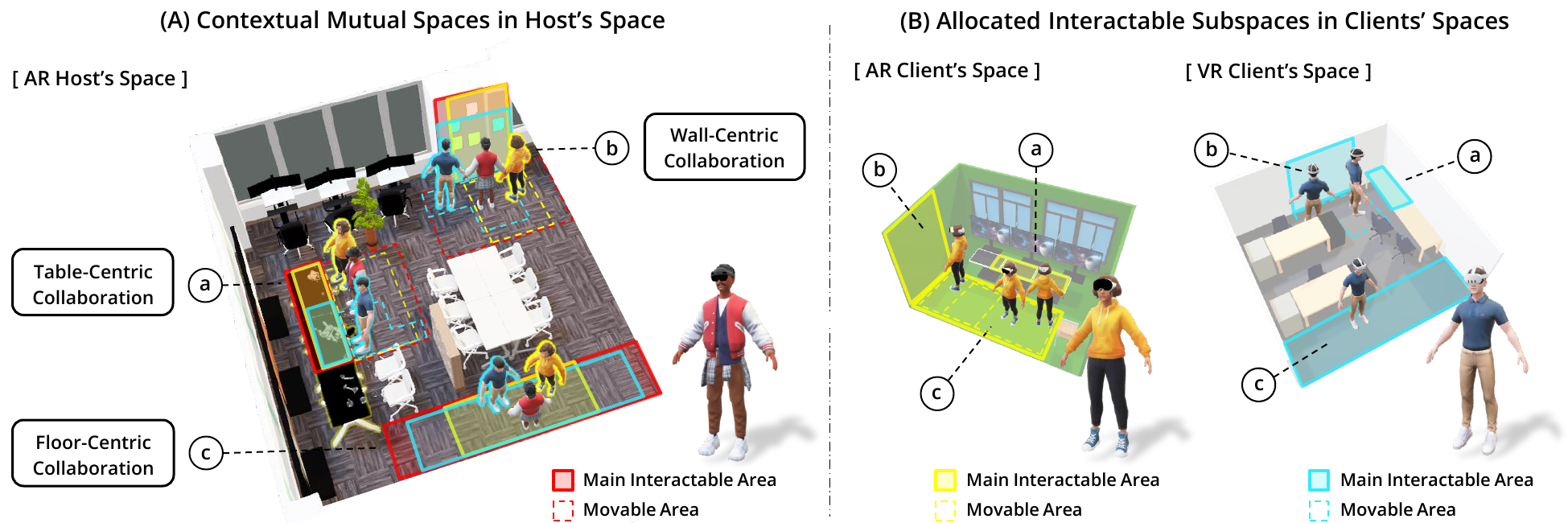}
  \caption{(A) All users can see the same AR host's space but have different interactable subspaces for robust mutual space generation as the number of remote clients increases: AR host (red), AR client (yellow), and VR client (blue). (B) Remote AR/VR clients access the host's space through their avatar, and their interactable subspaces and positions are determined in each context.
  }
  \label{fig:teaser}
}

\abstract{To enable remote Virtual Reality (VR) and Augmented Reality (AR) clients to collaborate as if they were in the same space during Mixed Reality (MR) telepresence, it is essential to overcome spatial heterogeneity and generate a unified shared collaborative environment by integrating remote spaces into a target host space. Especially when multiple remote users connect, a large shared space is necessary for people to maintain their personal space while collaborating, but the existing simple intersection method leads to the creation of narrow shared spaces as the number of remote spaces increases. To robustly align to the host space even as the number of remote spaces increases, we propose a spatial affordance-aware interactable subspace allocation algorithm. The key concept of our approach is to consider the perceivable and interactable areas separately, where every user views the same mutual space, but each remote user has a different interactable subspace, considering their location and spatial affordance. We conducted an evaluation with 900 space combinations, varying the number of remote spaces as two, four, and six, and results show our method outperformed in securing wide interactable mutual space and instantiating users compared to the other spatial matching methods. Our work enables multiple clients from diverse remote locations to access the AR host's space, allowing them to interact directly with the table, wall, or floor by aligning their physical subspaces within a connected mutual space.
} 


\begin{document}


\firstsection{Introduction}

\maketitle

Amidst the COVID-19 pandemic, numerous companies have adopted remote collaboration, reaping various positive outcomes such as flexibility, convenience, and cost savings. However, remote collaboration with limited interaction, such as 2D video conferencing, has resulted in low levels of presence and coexistence. From this perspective, a Mixed Reality (MR) telepresence system where remote clients access the Augmented Reality (AR) host's space through AR/MR/Virtual Reality (VR) Head Mounted Displays (HMDs) is gaining attention as the next era of remote collaboration to enable co-presence between remote users~\cite{piumsomboon2018mini, shin2022collaboration, yoon2023effects, teo2019mixed}. Since multiple users were situated in heterogeneous environments, one of the critical issues is overcoming spatial dissimilarity and generating a mutual space where remote users could interact with objects in the host's space~\cite{hoffman1998physically, kim2023edge}. However, manual adjustment of physical space to match the AR host's space is costly and even more problematic for collaboration across highly dissimilar and multiple spaces. To handle these issues, the MR telepresence system should understand the spatial configuration of host and remote spaces and figure out optimal aligning areas for each remote client's space.

For immersive MR remote collaboration, it is important that each user can intuitively interact with the table, wall, or floor in their connected virtual space by registering their physical space to the virtual space~\cite{kim2023edge, hoffman1998physically, gronbaek2023partially}. Recent work of mutual scene synthesis generates a virtual scene corresponding to the semantic features of multiple dissimilar physical spaces, and connected users could interact with semantically aligned areas~\cite{keshavarzi2022mutual, keshavarzi2020optimization}. However, there needs to be more evaluation of matching three or more remote spaces. On the other hand, the surface-centric alignment was more robust to spatial dissimilarity~\cite{gronbaek2023partially}, but it did not consider the precise alignment of surfaces with different scales. Another problem with MR telepresence situations is that when a remote user's avatar instantiates in another user's space, the immersion is broken when the remote avatar overlaps the on-site user~\cite{gronbaek2023partially}. To this end, a new spatial matching method is needed that can robustly match interactable areas even if multiple users are accessing from dissimilar spaces, and that takes into account the instantiation location of each remote user's avatar to avoid them overlapping on-site users.

We propose a spatial affordance-aware interactable subspace allocation (SA-ISA) method that allows AR/VR clients from multiple remote spaces to access the AR host's space by allocating different interactable subspaces to each remote user while all users see the entire AR host's space. \cref{fig:teaser} shows this concept, and this idea is motivated by real-world collaboration in which people physically interact with their surroundings while they can see all space. Considering the affordances of the space, we focused on three major collaboration contexts with general interaction targets: table, wall, and floor. We assumed the AR host user selected the collaboration context from the above three interaction targets, and the remote clients from multiple remote spaces could access and interact in the host's space through the following four processes. The first step is to find the optimal physical objects of each client space with the selected target from the host space using the scene graph. The second step is to find optimal positions for each client space with an optimization algorithm to secure a wide semantically aligned interactable area, primarily considering the main interactable area. The third step is determining the optimal instantiation location of clients and hosts based on the aligned interactable areas to allow utilization of their own physical geometry while preventing invading each user's personal space. The last step is subspace extraction from each client's space to ensure that each client has sufficient interactable space while minimizing semantically unmatched areas.

To evaluate the superiority of the proposed SA-ISA, we experimented with three main interaction targets (table, wall, and floor) and named each method SA-Table, SA-Wall, and SA-Floor. We compared our method with the semantic total intersection (S-TI) method, which defines a mutual space as the total intersected area~\cite{keshavarzi2020optimization}, and the semantic interactable subspace allocation (S-ISA) method, which only set a maximizing semantic match ratio without considering the collaboration context and interaction target. For evaluation, we combined four host spaces and ten client spaces, varying the number of client spaces to two, four, and six (900 space combinations). First of all, the user instantiation success rate of subspace-based matching methods (SA-ISA and S-ISA) succeeded in instantiating users in most cases even as the number of clients increased if there were physically instantiable areas in each context, while S-TI could only place 13.25\% with one host and four client spaces and could not allocate users in spatial combination with six client spaces. Moreover, results show that the SA-ISA could secure wider total interactable space according to the increase of client spaces, while S-TI's total interactable space decreases. We also analyzed the spatial alignment results qualitatively. We concluded that SA-ISA effectively creates mutual space by reflecting on the collaboration context, while S-ISA and S-IT only apply to floor-centric standing scenarios.

The contributions of this work can be threefold. First, we proposed a novel spatial affordance-aware interactable subspace allocation algorithm for an MR telepresence system that could be used for AR/VR clients from multiple remote spaces to access the AR host's space. Second, we introduced an optimal user instantiation algorithm that considers the spatial affordance and the proxemics between users to prevent remote avatars from invading other users' personal spaces. Finally, we evaluated that SA-ISA (ours) can generate collaborative spaces that are not only quantitatively superior in securing mutual space for interaction and user instantiation success rate but also qualitatively appropriate to the collaboration context through evaluation with 900 realistic space combinations. By utilizing SA-ISA with a playground boundary system used in typical VR HMDs, developers could summon multiple remote clients in proper locations to interact with nearby interaction targets by registering each client's physical space and connecting the AR host's space.

\section{Background}

\subsection{Dissimilar Space Matching}

Since the AR host's space accessed through AR/VR HMDs looks different from the physical spaces where remote users are located, it is necessary to match the disparate spaces for collaboration with co-presence. The simplest way to do this is finding a mutual walkable area by simply intersecting or aligning floor areas between spaces~\cite{lehment2014creating, sousa2016remote}. To allow users to move around a larger-than-real VR space, several studies have been proposed to apply Redirected Walking (RDW) within the user's perceptual threshold~\cite{razzaque2005redirected, kim2022configuration, steinicke2009estimation}. Although RDW can be utilized to allow users to move around a larger virtual space, it has the limitation that it can only match the movable area because the physical and virtual spaces are distorted by the redirection gain~\cite{neth2012velocity, nescher2016simultaneous, li2021openrdw}. RDW controllers for multiple VR users beyond a single VR user could adaptively redirect users considering the trajectories of multiple users~\cite{xu2023multi, bachmann2019multi, stormer2023study}, and other RDW controllers have divided physical and virtual spaces into polygons and applied appropriate redirection gains for each polygon~\cite{williams2021arc, williams2021redirected}. However, most of the RDW controllers have the limitation that the user may have an uncomfortable experience when applying the redirection gain by moving between the boundary values of the threshold to maximize the redirection effect~\cite{sakono2021redirected}.

On the other hand, a scene graph, a data structure that utilizes nodes and edges to depict the properties and relationships of objects~\cite{armeni20193d, hughes2022hydra}, can be employed to detect object clusters from dissimilar spaces and generate new scenes by accounting for the relationships between objects~\cite{fu2017adaptive, zhou2019scenegraphnet}. Jo et al.~\cite{jo2015spacetime} correlated two conference rooms with different arrangements of chairs for AR telepresence using scene graphs to determine the optimal position of each use, which is robust to slight scale and position differences. Furthermore, Keshavarzi et al.~\cite{keshavarzi2022mutual} utilize the scene graph from semantically intersected mutual space to create a new collaborative scene. Likewise, a scene graph can be used to find proper object cluster sets from dissimilar spaces by utilizing the semantic information of the space along with the relational meaning between objects~\cite{kim2024object}. Although it is effective in measuring the relational similarity of objects through scene graph comparison, there is a lack of research that actively uses scene graphs for spatial matching. 


\subsection{Mutual Space Generation and User Positioning}

Beyond matching two different spaces, several studies aim to create an MR mutual space to allow remote AR/VR users from dissimilar physical spaces to access the target collaboration space. The most common method involves using semantically labeled spatial data to maximize the movable area or set intersected areas for semantic matching, but this often results in narrow mutual spaces and neglects collaboration context and user-space interactivity~\cite{lehment2014creating, keshavarzi2020optimization}. On the other hand, a method for transforming and matching spaces using translation gain-based RDW has been proposed that considers the interactivity of users and spaces~\cite{kim2021adjusting, kim2022mutual}. For precise matching with objects that the user interacts with, quantitative indicators to measure interactable boundary and plane were proposed~\cite{kim2023edge}, but it has the limitation that it can only be used for VR users. Numerous attempts have aligned dissimilar spaces around a main target object, such as a telepresence system that projects a remote user onto a sofa for seated collaboration~\cite{pejsa2016room2room}. Another approach involves modular space connection, extending the area around walls to scale multiple different spaces easily, but it has the limitation of requiring additional equipment~\cite{zhang2022virtualcube, lawrence2021project}.


\begin{figure*}[ht]
 \centering
 \includegraphics[width=\linewidth]{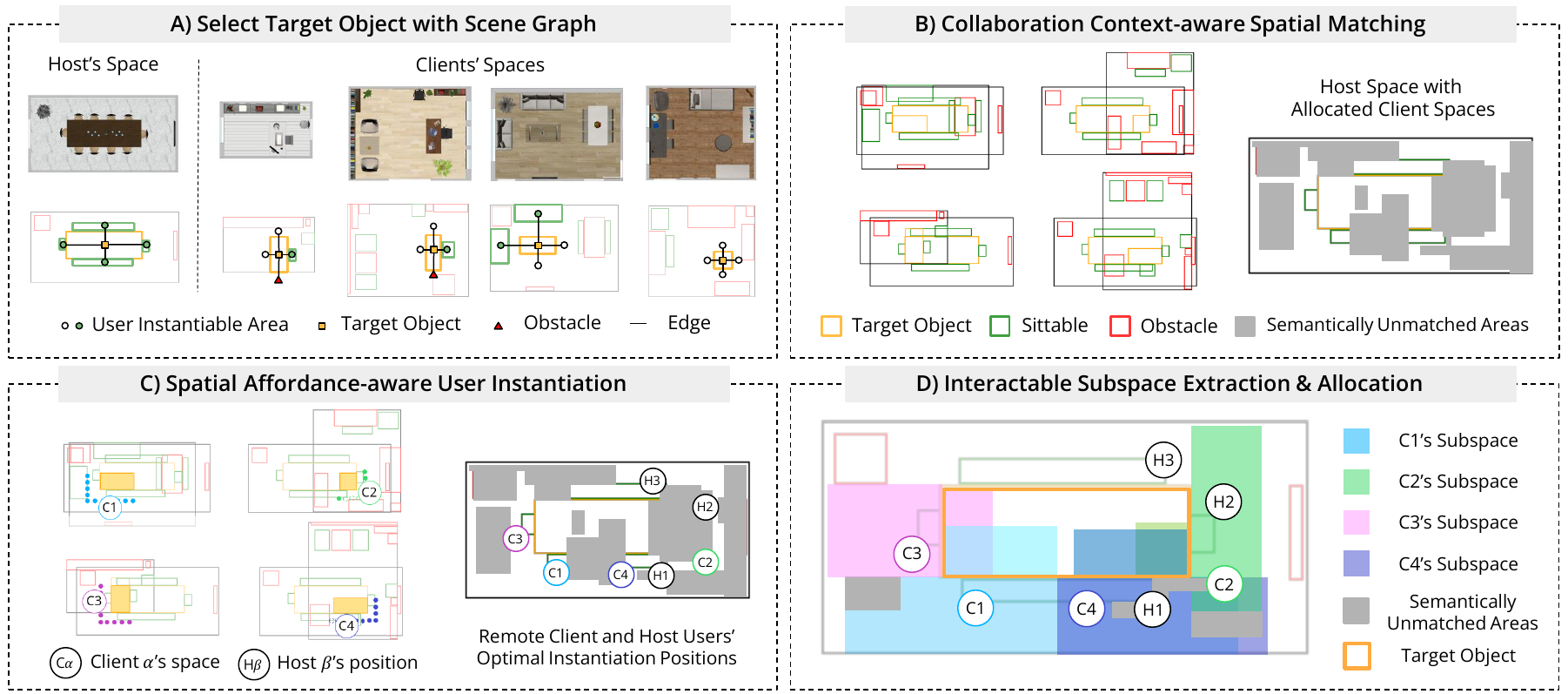}
 \caption{The overview of spatial matching for MR mutual space generation in a table-centric collaboration. A) Select a target object from each client's space corresponding to the host's target object with a scene graph, B) collaboration context-aware spatial matching, C) spatial affordance-aware user instantiation, and D) interactable subspace extraction and allocation.}
 \label{fig:spatial-matching-overview}
\end{figure*}

In MR telepresence with 3D avatars, adaptive placement of remote users in their own space and determining their position in the connected host's space should be considered. Yoon et al.~\cite{yoon2020placement, yoon2021full} attempted to correlate users' physical positions with remote avatars' positions to manage large spatial heterogeneity, but sudden teleportation of remote avatars due to discontinuous spatial coordinates disrupted immersion. Another study proposed to predict and reposition the avatar's trajectory considering the interacting object~\cite{wang2022predict}. It works when the next target object is clear; otherwise, it can lead to awkward turns if the prediction fails. In most MR telepresence situations, where remote users' physical spaces are smaller than the AR host's space, necessitating a visual transition to the interaction area and large-scale movement of the avatar~\cite{rhee2020augmented}, issues can arise, such as invading other users' personal space or augmented avatars overlapping with other users~\cite{gronbaek2023partially}. Therefore, when multiple remote users access an AR host's space, it is important to generate a mutual space by matching each remote space to the host's space and adaptively positioning remote avatars.


\section{Methodology}

In this study, we proposed an interactable subspace allocation method and optimal user instantiating algorithm for generating mutual space from multiple spaces. Our target scenario is for AR/VR clients from multiple remote spaces to have access to the AR host's space. This research aims to propose a spatial matching method that could generate a mutual space even if clients from four or more dissimilar remote spaces have access to the AR host's space. The idea comes from a real-world meeting scenario with more than six people at a big table; the area that each user interacts with stays around where they are located. So even the remote users could view the same perceivable area of the AR host, we focused on precisely tailoring each user's nearby interactable area. In other words, we assigned different interactable areas to each remote client. In this study, an interactable area is an area that users can touch with their hands or walk around. The allocated subspace refers to the partial area of each client's physical space registered in the AR host's space, which the client user can interact with their hands or physically walk around.

\cref{fig:spatial-matching-overview} shows the four sequential steps of our method. First, we assumed the host selects the collaboration context from three major interaction targets (table, wall, and floor), and our system will distinguish the most suitable objects in each client space with the scene graph. Second, receiving semantically labeled floorplans of host and client spaces as inputs, our optimization algorithm figures out the optimal position of each client's space to the AR host's space. Thirdly, our user instantiation algorithm will find the optimal instantiation positions of users considering spatial affordance and proxemics. Finally, interactable subspace for each client space is extracted and allocated based on the selected client's position. The initial collaboration environment can be set up through a series of steps by spatially matching the collaboration context, defining mutual spaces, and instantiating remote avatars so they do not invade other users' personal spaces.

\subsection{Target Object Selection with Scene Graph}
We start with the collaboration scenario in which the AR host organizing the MR remote collaboration has selected the target object to collaborate on in their space. Since there are various 3D model generation methods, such as Apple's Roomplan\footnote{https://developer.apple.com/augmented-reality/roomplan/}, with automatic semantic labeling from the scanned room ~\cite{armeni20163d, qi2017pointnet, avetisyan2024scenescript}, and by projecting these semantic 3D models onto a floor surface, it can be directly applied to our algorithm. Therefore, we assumed that pre-processed spatial information of physical space is already given and focused on the stages of the spatial matching algorithm in this study. The semantic information of the scene graph considers four different categories of affordance objects: table, wall, floor, and chair. All objects except the table and chair were tagged as obstacles. From receiving the semantically labeled floorplans of host and client spaces as inputs, target object selection is performed differently depending on whether the collaboration context requires an object-centered collaboration space. In this process, a scene graph based on the geometric and semantic information of objects is utilized to match the target object most suitable for each context~\cite{armeni20193d, wang2019planit, hughes2022hydra, kim2024object}. The object's geometric information refers to the size and location information of the bounding box. 


\cref{fig:spatial-matching-overview}A) shows a target object selection with a scene graph in table-centric collaboration. In this case, the scene graph was formed centered on the table according to the affordance and direction of directly related objects. The nodes connected to the central node (the table) include an obstacle object and a chair object with sittable affordance. The edge of the scene graph is formed when an obstacle or sittable object is around the table, considering a user's personal area size. Considering a human's proxemics and remote proxemics~\cite{hall1966hidden, sousa2016remote} of intimate space $0.45 m$ with $0.15 m$ margin, we set this personal area size as $0.6 m$ in this study. Then, we scored a matching rate for all tables in the client, choosing the table with the highest score as the target interactable object. Higher scores are assigned when similar affordance objects are in the same direction~\cite{kim2024object}. The above matching process is performed for each table with a 90-degree rotation in each space, and preprocessed space can be obtained using the table with the highest matching rate and the rotation angle. Similarly, in the case of wall-centric collaboration, the wall with an adjacent wide floor was selected as the target object. In the case of floor-centric collaboration, the movable floor is the target object. By understanding the relationships between objects through a scene graph and considering the semantic area relevant to the collaboration context, optimal object clusters and the rotation values for each space can be obtained.

\subsection{Collaboration Context-aware Spatial Matching} 

Through the above process, we obtained floorplans of each space and the target object to be considered primarily in the spatial matching process. In this study, we divided the collaboration context into three main types and performed spatial matching accordingly. The first collaboration context is a table-centric scenario where users sit around a table and communicate with each other. \cref{fig:spatial-matching-overview}B) show the example of a table-centric scenario and corresponding spatial matching results that each client space's table boundary and host's table boundary were aligned. We named this table-centric spatial matching as SA-Table. The second is vertical surface-based collaboration. In this case, a mutual space is created for scenarios where users collaborate by placing post-its on the wall or having a meeting by writing on the board. We named it as SA-Wall. The last collaboration context is where users stand around talking in an empty space. We named this type as SA-Floor. The objective function to perform spatial matching according to these three representative collaboration contexts can be formalized as follows:

\begin{align} 
    O(\phi) = & \omega_{1}\psi_{G, sem}+\omega_{2}\psi_{G, size} \nonumber \\ 
    & + \omega_{3}\psi_{I, hor} + \omega_{4}\psi_{I, ver} + \omega_{5}\psi_{I, mov}.
\label{equ:objFunc}
\end{align} 

\cref{equ:objFunc} consists of two main components: the geometric term ($\psi_{G}$) and the interaction term ($\psi_{I}$). The geometric term contains a semantic match ratio ($\psi_{G, sem}$) and a matched space size ratio ($\psi_{G, size}$), which are general metrics that have been utilized in previous spatial matching methods~\cite{keshavarzi2020optimization, lehment2014creating}. These components are generally utilized when matching heterogeneous spaces from a spatial perspective, aiming to create a large collaborative space from different spaces. Next, the interaction term consists of three metrics considering users' interactions, such as touch or walking~\cite{kim2023edge}. First is the horizontal boundary sync ratio ($\psi_{I, hor}$), which measures how well the boundaries of the horizontal surface match. The second metric vertical surface sync ratio ($\psi_{I, ver}$) measures how well the vertical surface being interacted with in the virtual space is aligned with the vertical surface in the user's physical space. The final metric is the movable floor sync ratio ($\psi_{I, mov}$), which measures the aligned movable floor between host and client spaces. For SA-Table, only $\psi_{I, hor}$ is utilized, and the other two interaction terms are not used. Similarly, for SA-Wall, only $\psi_{I, ver}$ is used as an interaction term, and for SA-Floor, only ($\psi_{I, mov}$) is used for spatial matching. By performing the optimization algorithm using the objective function, we could achieve space matching for each host space and client space pair according to the collaboration context.

\subsection{Spatial Affordance-aware User Instantiation}

\begin{figure*}[ht]
 \centering
 \includegraphics[width=\linewidth]{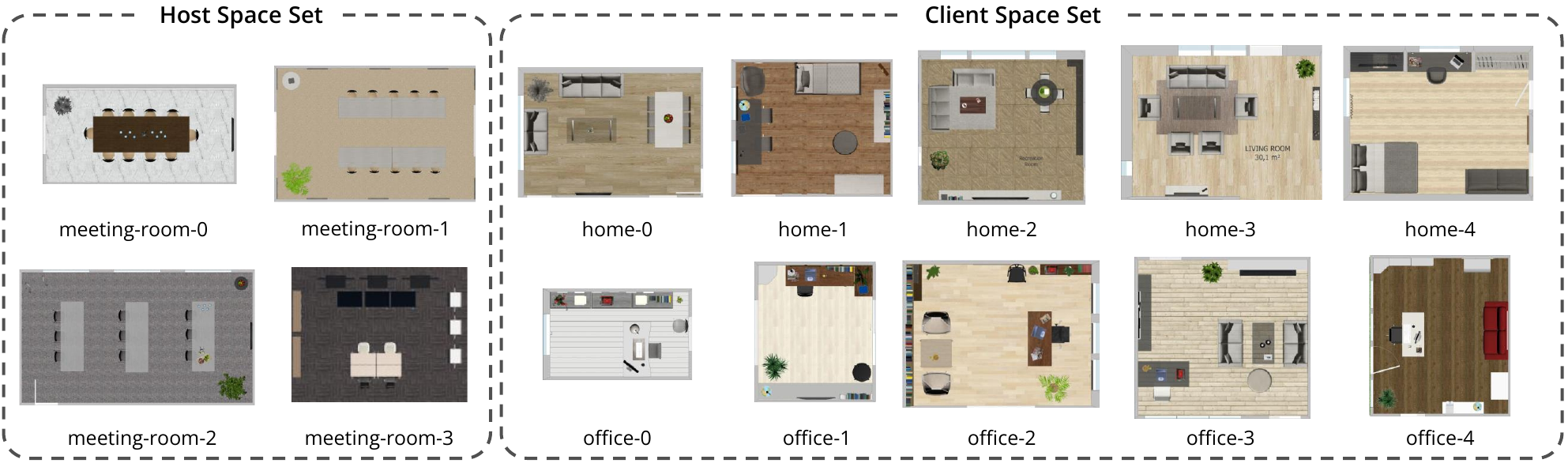}
 \caption{Top view of four host spaces (four meeting rooms) and ten client spaces (five homes and five offices) used for evaluation.}
 \label{fig:input-spaces}
\end{figure*}

Based on the optimal spatial matching, we propose a spatial affordance-aware user instantiation algorithm that finds the optimal starting position for users to prevent break-of-immersion by remote users' avatars invading other user's personal space~\cite{gronbaek2023partially, von2019you}. \cref{fig:spatial-matching-overview}C) shows the optimal instantiation position of each user in a table-centric collaboration scenario. Each client and host are located in corresponding positions considering the spatial affordance and personal area. Before beginning the collaboration, remote clients will be guided to the optimal initial position in their physical space and summoned to the host's mutual space, considering the relative position decided from the spatial matching results.

We extracted the possible position candidates considering the collaboration context to find the optimal starting position set. For a table-centric scenario, we sampled the user's position with step size ($d_{step} m$) around a matched boundary. We set the user's center to be $0.45 m$ away from the horizontal surface's boundary, considering a circle with a diameter of minimum personal area ($0.6 m$). Similarly, we sampled the candidate positions with $d_{step} m$ for wall-centric collaboration, separating $0.45 m$ from the matched perpendicular wall. Finally, for the floor, the user's position is sampled on the floor with a sampling interval of $d_{step} m$. Then, we exclude position candidates where users could not instantiate, such as the table, wall, obstacle, and semantically unmatched areas. We remained a sittable label as user instantiable areas for SA-table since collaboration around a table often involves sitting down. Furthermore, we eliminated position candidates when the user's personal space ($0.6 m$) was not secured. Selected position candidates were then combined to find the conditions under which the clients could be positioned so that the gap between them was at least $0.6 m$ and the target distance was the closest. The user instantiation algorithm first reduces the position candidates by increasing personal space and then finding the optimal users' positions.

\subsection{Interactable Subspace Extraction \&  Allocation}

Based on each client's instantiation position, we propose an interactable subspace extraction and allocation method to reduce the semantically unmatched area while ensuring the interactable space for users. \cref{fig:spatial-matching-overview}D) shows the subspace extraction and allocation result in the table-centric scenario, and it shows much fewer semantically unmatched areas compared to \cref{fig:spatial-matching-overview}B)'s host space with allocated client spaces. To do so, we first create four markers with $l_{m} m \times 0.1 m$ and the $l_{m}$ side of the marker parallel to the square that makes up the user's personal area at $0.3 m$ from the user's optimal position. Then, move four markers on each side up, down, left, and right by $0.1 m$ until they collide with the semantically unmatched areas. We set initial $l_{m}$ as $0.6 m$, the minimum personal space, which goes larger depending on the detection thresholds. If the system determines that the user has enough interactable area, subspace extraction focuses more on minimizing the semantic unmatched area by increasing the marker's width. This allows us to extract a subspace from each client's space that minimizes unnecessary object augmentation while securing their interactive space.


By allocating clients' subspaces to the host space, each client has a different interactable subspace, as shown in \cref{fig:spatial-matching-overview}D). The final mutual space can be created by drawing a wall across considered allocated subspaces' boundaries. This simple form of diminished reality creates a mutual space, and the semantically unmatched parts are augmented in the host space. Through this subspace allocation approach, we could ensure a wide interactable area even if the number of client spaces increases by separating the perceivable and interactable areas. Each client can only utilize the subspace allocated to them, and it is necessary to ensure that each remote user does not cross the boundary through safety zone visualization\footnote{https://www.meta.com/help/quest/articles/in-vr-experiences/oculus-features/boundary/}.


\section{Evaluation}

\subsection{Study Design and Quantitative Metrics}

We set two comparison groups to evaluate our spatial affordance-aware interactable subspace allocation (SA-ISA) method. The first comparison method is a semantic interactable subspace allocation (S-ISA), which adapts the subspace allocation method we have developed but only considers geometric components. The second comparison method is a semantic total intersection method (S-TI), which only considers geometric terms and defines the mutual space as the total intersection area of all given spaces. For the quantitative evaluation, we utilized five metrics: one is the user instantiation success rate, and four metrics related to the matched space's area. At first, we determine whether all users can be placed while securing minimum personal space for each user. Considering computation time and a sufficient number of possible position candidates, we set $d_{step}$ for sampling each user's position with a step size of $0.2 m$. The user instantiation success rate is defined as the ratio of a number of space combinations where success in positioning the user by the total number of space combinations, which also allows us to determine whether mutual space can be created.

Next, we utilized four metrics related to the interactable and obstacle areas of the matched space. In order to focus on the spatial matching results, we utilized the spatial matching results when the number of host users is one. First, we used the mean total interactable area, the average of the total union of all possible client areas. The interactable area means the client's space was aligned with one semantic label with the host's space so that they can interact with their physical space. In addition, we defined the mean total obstacle area as the sum of matched obstacle areas and semantically unmatched areas. A large mean total obstacle area indicates that many unnecessary objects should be augmented, so a smaller value means good spatial matching. The remaining two metrics are the mean interactable subspace area per client and the mean obstacle area per client. We decided to utilize these two metrics because SA-ISA allocates different interactable areas corresponding to each client. Whereas in the case of the S-TI, the total interactable area is the same as the interactable subspace area per client. By measuring both interactable and obstacle areas separately, we could compare the interactable area for each client according to spatial matching methods and the obstacle area for each client to see how much unnecessary object augmentation was needed.

\begin{figure*}[ht]
 \centering
 \includegraphics[width=\linewidth]{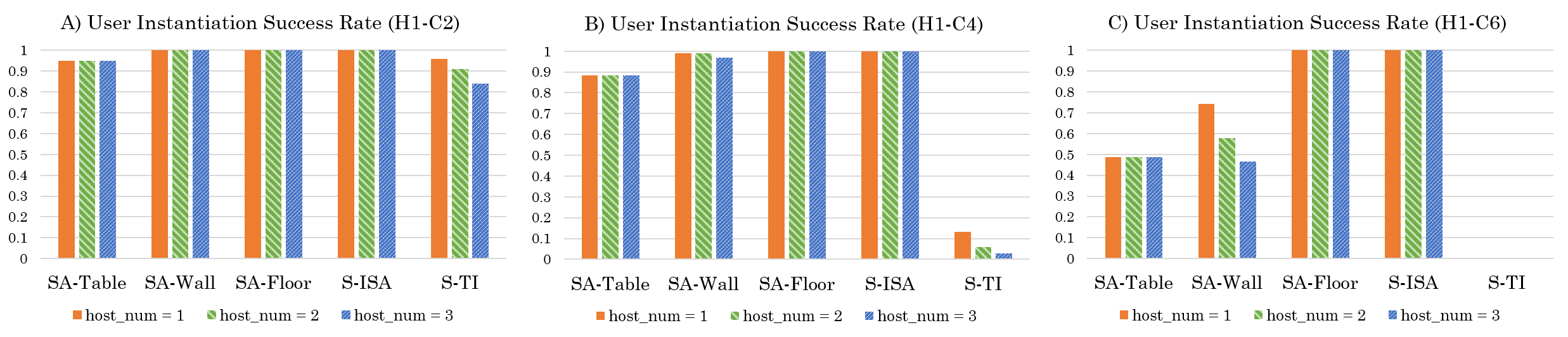}
 \caption{The user instantiation success rate in each condition. A) H1-C2, B) H1-C4, and C) H1-C6.}
 \label{fig:success-rate}
\end{figure*}

\begin{figure*}[ht]
 \centering
 \includegraphics[width=\linewidth]{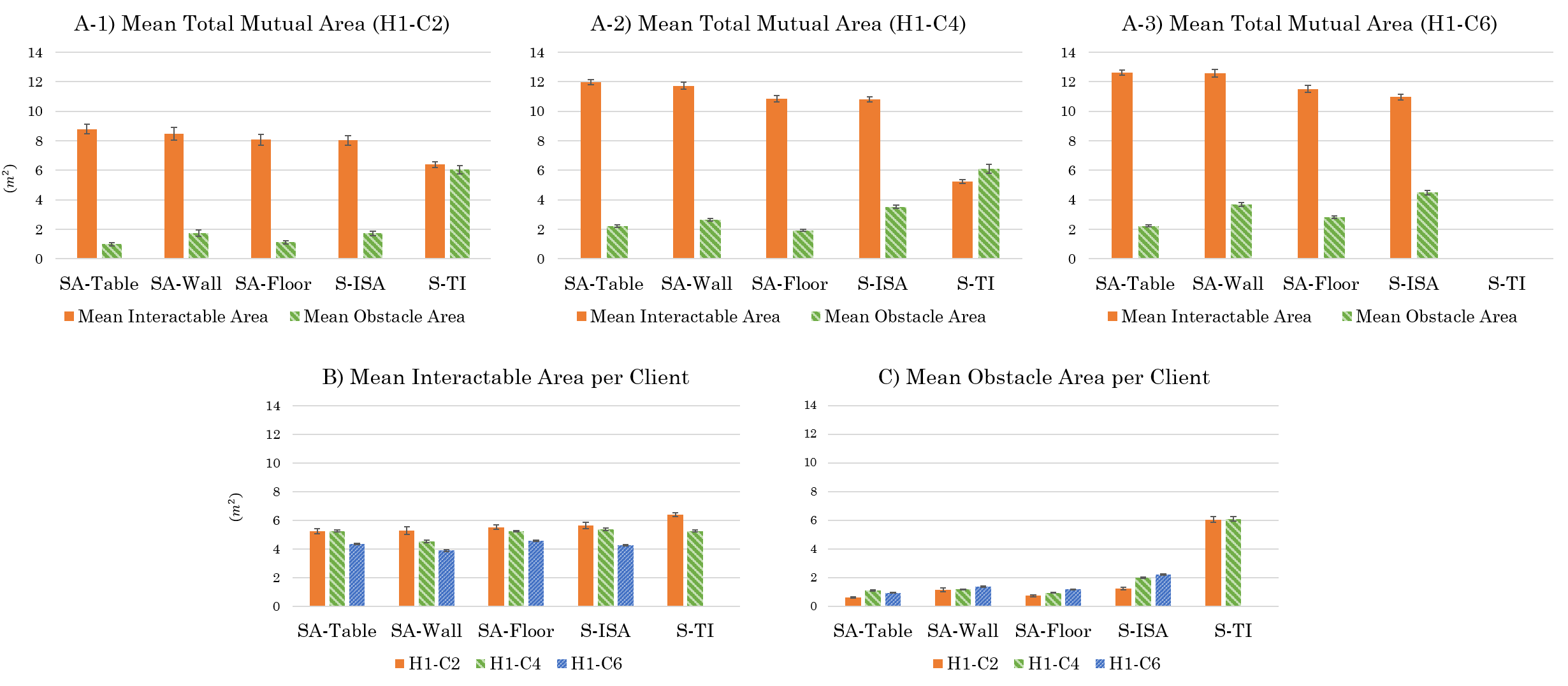}
 \caption{The mean of interactable space area and obstacle area in each condition. A) Mean total mutual space area in A-1) H1-C2, A-2) H1-C4, and A-3) H1-C6. B) Mean interactable area per client and C) mean obstacle area per client.}
 \label{fig:mean-area-result}
\end{figure*}

\subsection{Input Data and System Implementation}

In this study, we utilized four host spaces and ten client spaces. \cref{fig:input-spaces} shows 14 input spaces created by floor planning based on real spaces. We assumed each space has a semantic label and starts from receiving semantically labeled floorplans of each space as inputs. There were four host spaces consisting of three typical meeting rooms and one specialized room setup for XR telepresence\footnote{https://www.cisco.com/c/en/us/products/collaboration-endpoints/immersive-telePresence/}. For client spaces, five rooms from homes and five office spaces were selected as typical spaces for remote users. Since we varied the number of client spaces, H1-C2 refers to a host space with two client spaces, H1-C4 refers to a host space with four client spaces, and H1-C6 refers to a host space with six client spaces. For H1-C2 condition, we selected one space from host spaces, one from homes, and one from the offices, creating $\sideset{_4}{_1}{\operatorname{\text{C}}} \times \sideset{_5}{_1}{\operatorname{\text{C}}} \times \sideset{_5}{_1}{\operatorname{\text{C}}} = 100$ space combinations. Similarly, for space combinations with H1-C4, we selected one from host spaces, two from home, and two from the office, creating $\sideset{_4}{_1}{\operatorname{\text{C}}} \times \sideset{_5}{_2}{\operatorname{\text{C}}} \times \sideset{_5}{_2}{\operatorname{\text{C}}} = 400$ space combinations. Finally, for H1-C6, one from host spaces, three spaces from home, and three from the office were selected, for a total of $\sideset{_4}{_1}{\operatorname{\text{C}}} \times \sideset{_5}{_3}{\operatorname{\text{C}}} \times \sideset{_5}{_3}{\operatorname{\text{C}}} = 400$ space combinations. To this end, we experimented with a total of 900 space combinations. 

For optimization, the weight for geometric terms, $\psi_{G, sem}$ and $\psi_{G, size}$ in ~\cref{equ:objFunc}, were set as ten for every spatial matching method. In the case of S-IT and S-ISA, three interaction terms were all set as zero. For SA-ISA (SA-Table, SA-Wall, and SA-Floor), the weight of each interaction term was set as 100 for the corresponding context, and other interaction terms that are not relevant to the target context were set as zero. For instance, each weight in the ~\cref{equ:objFunc} for SA-Table was set as $\omega_{1} = 10, \omega_{2} = 10, \omega_{3} = 100, \omega_{4} = 0,$ and $\omega_{5} = 0$. In this experiment, we assumed a single interaction target situation. However, we have built the algorithm so that if multiple interaction targets are all important, the weights can be adjusted to optimize for each scenario. We conducted our evaluations on a desktop with an Intel i9-10900X CPU, NVIDIA RTX 3090 GPU, 64GB RAM, and 2TB SSD. For the optimization environment, we utilized meta-optimization algorithm from Pygmo 2.19.0~\cite{Biscani2020}, based on self-adaptive fitness formulation by Farmani et al.~\cite{farmani2003self}, in the Python 3.9.13 Anaconda virtual environment~\cite{anaconda}. Moreover, we used Pypex\footnote{https://github.com/mikecokina/pypex/} for computation with polygons. Python's Pillow (PIL) 9.4.0 library~\cite{umesh2012image} was used for plots.

\subsection{Results}

\subsubsection{User Instantiation Success Rate}

\cref{fig:success-rate} shows the user instantiation success rate of SA-ISA (SA-Table, SA-Wall, and SA-Floor) and two comparison groups (S-ISA and S-TI). \cref{fig:success-rate}A) shows the success rate for 100 space combinations for H1-C2. Results show that SA-Table has a success rate of 0.95, and SA-Wall and SA-Floor succeed in allocating users in every condition. S-ISA could instantiate users in all cases regardless of the number of hosts. Lastly, S-TI had a success rate of 0.96 when the number of hosts was one, 0.91 with two, and 0.84 with three hosts. \cref{fig:success-rate}B) shows the user instantiation success rate for H1-C4's 400 space combinations. SA-Table had a success rate of 0.8825 for all, while SA-Wall shows 0.99 with one or two hosts and 0.97 with three hosts. Moreover, SA-Floor and S-ISA were successful in every condition. On the other hand, S-TI's user instantiation success rate was 0.1325 when there was a single host. When the number of hosts was two, S-TI's success rate was 0.06, and it was 0.0275 with three hosts. Finally, \cref{fig:success-rate}C) represents the user instantiation success rate for H1-C6's 400 space combinations. SA-Table's success rate was 0.4875, regardless of the number of hosts. SA-Wall had a success rate of 0.7425 with one host, 0.5775 with two hosts, and 0.465 with three host users. SA-Floor and S-ISA succeeded in placing users in all cases. Lastly, S-TI failed to instantiate users in every condition in H1-C6.

\begin{figure*}[ht]
 \centering
 \includegraphics[width=\linewidth]{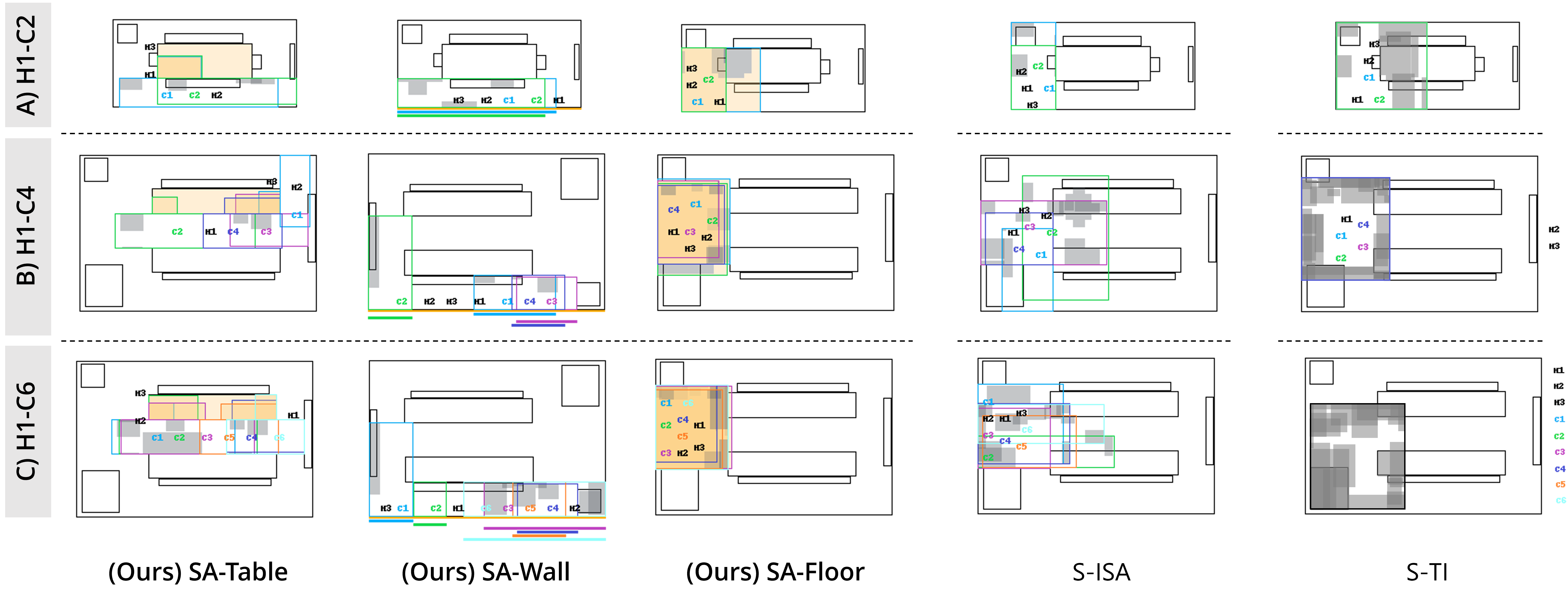}
 \caption{The representative spatial matching results of our method (SA-Table, SA-Wall, and SA-Floor) and comparison groups (S-ISA and S-TI). A) H1-C2, B) H1-C4, and C) H1-C6. H1, H2, and H3 refer to the sequential order of instantiated host users. C1 from C6 refers to clients from each space. The gray areas refer to semantically unmatched areas, and the instantiation failed users were written outside the host's space.}
 \label{fig:spatial-matching-result}
\end{figure*}

\subsubsection{Mean Total Mutual Space Area}

\cref{fig:mean-area-result}A) shows the plots of the mean total interactable area and mean total obstacle area in the corresponding conditions with error bars using the standard error of the mean (SEM). \cref{fig:mean-area-result}A-1) shows the matching results of H1-C2. The mean total interactable area was 8.79 $m^{2}$ (SD = 3.09) for SA-table, 8.47 $m^{2}$ (SD = 4.47) for SA-wall, 8.08 $m^{2}$ (SD = 3.68) for SA-floor, 8.02 $m^{2}$ (SD = 3.34) for S-ISA, and 6.40 $m^{2}$ (SD = 1.94) for S-TI. The mean total obstacle area was 1.00 $m^{2}$ (SD = 0.96) for SA-table, 1.73 $m^{2}$ (SD = 2.18) for SA-wall, 1.12 $m^{2}$ (SD = 1.15) for SA-floor, 1.73 $m^{2}$ (SD = 1.44) for S-ISA, and 6.04 $m^{2}$ (SD = 2.85) for S-TI. \cref{fig:mean-area-result}A-2) shows the results of H1-C4. The mean total interactable area was 11.98 $m^{2}$ (SD = 3.07) for SA-table, 11.72 $m^{2}$ (SD = 4.54) for SA-wall, 10.84 $m^{2}$ (SD = 4.32) for SA-floor, 10.80 $m^{2}$ (SD = 3.58) for S-ISA, and 5.25 $m^{2}$ (SD = 0.85) for S-TI. The mean total obstacle area was 2.21 $m^{2}$ (SD = 1.77) for SA-table, 2.65 $m^{2}$ (SD = 2.09) for SA-wall, 1.92 $m^{2}$ (SD = 1.45) for SA-floor, 3.53 $m^{2}$ (SD = 2.37) for S-ISA, and 6.09 $m^{2}$ (SD = 2.19) for S-TI. \cref{fig:mean-area-result}A-3) shows the matching results of H1-C6. The mean total interactable area was 12.61 $m^{2}$ (SD = 2.32) for SA-table, 12.57 $m^{2}$ (SD = 4.60) for SA-wall, 11.51 $m^{2}$ (SD = 4.53) for SA-floor, and 10.96 $m^{2}$ (SD = 3.97) for S-ISA. The mean total obstacle area was 2.24 $m^{2}$ (SD = 0.76) for SA-table, 3.67 $m^{2}$ (SD = 2.25) for SA-wall, 2.82 $m^{2}$ (SD = 1.74) for SA-floor,and 4.48 $m^{2}$ (SD = 2.71) for S-ISA. In the case of S-TI, no spatial combination succeeded for user instantiation out of 400 combinations.

\subsubsection{Mean Allocated Subspace Area per Client}

\cref{fig:mean-area-result}B) shows the plots of the mean interactable area per client with error bars using the SEM. For SA-Table, it was 5.24 $m^{2}$ (SD = 2.44) at H1-C2, 5.24 $m^{2}$ (SD = 2.39) at H1-C4, and 4.34 $m^{2}$ (SD = 1.14) at H1-C6. For SA-Wall, it was 5.28 $m^{2}$ (SD = 3.72) at H1-C2, 4.52 $m^{2}$ (SD = 3.37) at H1-C4, and 3.89 $m^{2}$ (SD = 2.67) at H1-C6. For SA-Floor, it was 5.52 $m^{2}$ (SD = 2.27) at H1-C2, 5.24 $m^{2}$ (SD = 2.15) at H1-C4, and 4.58 $m^{2}$ (SD = 2.14) at H1-C6. For S-ISA, it was 5.64 $m^{2}$ (SD = 2.96) at H1-C2, 5.36 $m^{2}$ (SD = 2.99) at H1-C4, and 4.24 $m^{2}$ (SD = 2.28) at H1-C6. For S-TI, it was 6.40 $m^{2}$ (SD = 1.94) at H1-C2, 5.25 $m^{2}$ (SD = 0.85) at H1-C4. Since S-TI failed to instantiate users in H1-C6's every condition, the interactable area could not be created.

\cref{fig:mean-area-result}C) shows the plots of the mean obstacle area per client with error bars using the SEM. For SA-Table, the it was 0.61 $m^{2}$ (SD = 0.73) for H1-C2, 1.11 $m^{2}$ (SD = 1.36) for H1-C4, and 0.94 $m^{2}$ (SD = 0.77) for H1-C6. For SA-Wall, it was 1.15 $m^{2}$ (SD = 1.82) at H1-C2, 1.18 $m^{2}$ (SD = 1.44) at H1-C4, and 1.37 $m^{2}$ (SD = 1.45) at H1-C6. For SA-Floor, it was 0.74 $m^{2}$ (SD = 0.87) at H1-C2, 0.93 $m^{2}$ (SD = 0.95) at H1-C4, and 1.17 $m^{2}$ (SD = 1.17) at H1-C6. For S-ISA, it was 1.25 $m^{2}$ (SD = 1.24) at H1-C2, 2.01 $m^{2}$ (SD = 1.76) at H1-C4, and 2.21 $m^{2}$ (SD = 1.95) at H1-C6. In the case of S-TI, the results are the same as the above total because the interactable area of all clients is the same: 6.04 $m^{2}$ (SD = 2.85) at H1-C2, 6.09 $m^{2}$ (SD = 2.19) at H1-C4. Since S-TI failed to instantiate users in H1-C6, the obstacle area could not be measured.




\section{Discussion}

\subsection{Analysis}

The qualitative performance between SA-ISA and S-ISA can be found in \cref{fig:spatial-matching-result}, which is the representative spatial matching results with our method and two comparison groups (A: H1-C2, B: H1-C4, C: H1-C6). The gray areas in each spatial matching result indicate the semantically unmatched areas, and if the users are allocated on it, it refers to the sittable label. In the case of S-ISA, subspaces are allocated and matched to increase the semantic match ratio and create a large mutual space. Therefore, we observed that the spatial alignment results are similar to SA-Floor, which maximizes the floor that occupies the largest semantic label in a typical room. This means that S-ISA is mostly appropriate for floor-centric standing scenarios. On the other hand, \cref{fig:spatial-matching-result}'s SA-Table, SA-Wall, and SA-Floor show corresponding different spatial allocations considering the target interaction object. These spatial matching results support the idea that to create a mutual space for MR telepresence with interactable areas, collaboration-context and interaction terms must be considered rather than simply geometric terms.

As the number of remote spaces increased, the subspace allocation-based space matching methods had a higher user instantiation success rate than the S-TI. \cref{fig:success-rate}A) shows that when the number of client spaces is two, SA-Table shows a 95\% success rate regardless of the number of hosts, and SA-Wall, SA-Floor, and SA-ISA all succeed in user placement across 100 space combinations. In contrast, S-TI shows a decreasing trend in user instantiation success rate as the number of hosts increases, and this trend becomes more pronounced as the number of client spaces increases. \cref{fig:success-rate}B) show that when the number of client spaces increased to four, SA-Table had an 88.25\% success rate, SA-Wall had a 97\% success rate even with three host users, and SA-Floor and SA-ISA succeeded in placing users in all cases. However, the success rate of S-TI was 13.25\% with one host user, 6\% with two host users, and only 2.75\% with three host users. Finally, when the number of client spaces was six, the success rate was around 50\% for SA-Table, because the size of the collaboration tables used in meeting-room-2/-3 in \cref{fig:input-spaces} could not physically accommodate more than seven users. In the case of SA-Wall H1-C6, a user instantiation success rate of 74.25\% for one host user, 57.75\% for two host users, and 46.5\% for three host users. This is also due to physical constraints, such as meeting-room-3's wall length being unavailable for positioning seven people.

As shown in \cref{fig:spatial-matching-result}'s SA-Floor and S-ISA cases, two methods configured mutual space based on movable area, so there is relatively more space. Since the floor is the most dominant semantic information in most spaces, all users were successfully placed regardless of the number of hosts. Finally, S-TI failed to place users in all cases of H1-C6 conditions. This is because only the white area within the highlighted subspace can be utilized, as shown in the S-TI results in \cref{fig:spatial-matching-result}C). This intersected area also contains many small areas where users could not secure personal space, so it failed to instantiate at least seven users in H1-C6. Through these results, we show that the S-TI mostly fails to create mutual spaces for more than four client spaces. At the same time, our proposed subspace allocation-based approaches were more robust in creating mutual collaboration spaces regardless of the increase in the number of remote spaces.

\cref{fig:mean-area-result}A) show that the total interactable area of SA-ISA (SA-Table, SA-Wall, and SA-Floor) and S-ISA increased as the client space increased. When the number of remote spaces increases, the collaborative space should be larger because there are more users. Furthermore, \cref{fig:mean-area-result}B) shows that all subspace-based matching methods could secure more than $4 m^2$ of the interactive area except SA-Wall, which secured $3.89 m^2$ in H1-C6. This shows that SA-ISA not only secures an expansive total interactive space but also ensures that each client has an interactable area around their instantiation position. On the other hand, the S-TI method, which defines the intersection of all spaces as a shared collaboration space, fails to place users in most cases of H1-C4 and H1-C6, where more than four client spaces were matched. Even for the 13.25\% of spaces that succeeded with H1-C4, the total interactable area was only $5.25 m^2$, less than half of the $10.80 m^2$ area of the SA-ISA method, which had the smallest area among four subspace allocation methods. In the case of H1-C6, none of the 400 space combinations could create a shared space, confirming that the existing simple intersection method would be hard to support when summoning remote clients if the number of client spaces increased.

In addition, the subspace allocation method could not only maximize the interactive area but also reduce the unnecessary obstacle areas compared to the S-TI method. To prevent VR users' collision with physical objects and remote avatars passing through objects, immovable areas should be highlighted in areas with inconsistent information based on the spatial matching results. Larger unmatched areas require more immovable area visualization, so a mutual space with smaller unmatched areas is a better choice. \cref{fig:mean-area-result}A) shows most cases of SA-ISA had a mean obstacle area within $2 m^2$ except SA-Wall (H1-C6) was $3.67 m^2$. For SA-ISA, the total obstacle area was $3.53 m^2$ for H1-C4 and $4.48 m^2$ for H1-C6. On the other hand, in the mutual space generation succeeded case of S-TI, more than $6 m^2$ of obstacles were augmented in H1-C2 and H1-C4, respectively. This necessity to visualize immovable area is even more significant if we check the mean obstacle area per client in \cref{fig:mean-area-result}B). SA-ISA methods augmented around $1 m^2$ of obstacles per client, while S-ISA augmented around $2 m^2$ of area per client. In the case of S-TI, around $6 m^2$ of obstacle areas are not interactable. This confirms that SA-ISA can significantly reduce the visualization of unnecessary immovable areas through object augmentation compared to simple intersections.


\subsection{Limitations}

Although we presented a new method for registering multiple remote spaces to the AR host's space by allocating different subspaces to each client, several limitations remain. The first limitation is that our spatial allocation algorithm needs to consider group dynamics between users and is hard to perform in real-time. In this study, we focused on exploring how to mathematically allocate interactable subspaces so that each remote user could have an interactable area near their summoned location, but to apply it to real-world MR telepresence scenarios, it is necessary to adapt a feature that allows users to choose where to instantiate considering the relationships between participants. In addition, our current algorithm's computation time takes from a few seconds to a few minutes, depending on the complexity of the space and the number of spaces. Utilizing human-in-the-loop considering group dynamics and changing current CPU-based to GPU-based optimization will decrease computation time.

The second limitation is that SA-ISA does not take into account the vertical height difference between the host space and remote spaces. In this study, we assumed that the horizontal interactable planes in the host and client space have the same height to focus on the aligning boundary and interactable area of the horizontal plane. Also, aligning a wall does not account for vertical differences, assuming that the wall is flat without unevenness and high enough to be a typical room wall. However, in the real world, tables in each user's room have different heights, and in some cases, the wall may have layers or different heights. To overcome this limitation, it is necessary to utilize space inputs with more details, such as surfaces' height and shape characteristics, and adapt vertical redirection techniques for accurate 3D subspace allocation.

Finally, although the optimal subspace and user positions can be obtained, the effectiveness of subspace allocation has yet to be explored through user experiments. In particular, since there is a difference between the perceivable and interactable areas for each user, users may feel a breakpoint when they desire to interact with the area outside their registered subspace. Further user studies are needed to verify the usability and effectiveness of accessing the AR host's space through the subspace allocation-based method compared to the simple intersection method, where all clients have the same interactable area. Moreover, users' personal characteristics should be considered and inputted separately to perform spatial matching and user placement adaptively.


\section{Conclusion and Future Work}

This study proposed a novel interactable subspace allocation algorithm that considers spatial affordance for MR remote collaboration where remote AR/VR users from multiple dissimilar spaces access the host's space. Using a scene graph, we find the pair of interaction targets (table, wall, and floor) between host and client spaces and propose a spatial matching algorithm considering the collaboration context. It could secure interactable subspaces for each client from heterogeneous spaces while minimizing unnecessary obstacle augmentation. Furthermore, we developed a spatial affordance-aware user instantiation algorithm considering the proxemics. We evaluated our method with spatial matching experiments with 900 realistic spatial combinations, varying the number of client spaces to two, four, and six. Results show that the proposed SA-ISA could generate a larger total interactable space as the number of client spaces increases, and the unnecessary object augmentation is less than half of the simple semantical intersection method.

To address the aforementioned limitations, the algorithm could be improved by showing an interactive area that varies depending on where the user wants to be considering group dynamics or arranging people in a variety of facing formations depending on the task at hand~\cite{kendon1990conducting, marshall2011using}. Alternatively, converting the algorithm to GPU-based computation with a human-in-the-loop interface for real-time computation and considering group dynamics is also necessary. Next, avatar repositioning techniques should be combined to overcome the horizontal surface height differences between the host and the remote space. To overcome vertical differences, VR hand redirection~\cite{gonzalez2022model} or a deictic motion retargeting with a prediction mode can be used for both AR and VR clients~\cite{kang2023real}. Finally, it is necessary to create the final mutual space utilizing the subspace boundaries and initial positions of the users obtained from our algorithm. Within this generated mutual space, the user study considering individual characteristics is needed to validate the pros and cons of our subspace allocation-based approach. By enabling immersive MR remote collaboration in a generated mutual space, distant people can collaborate as if they were in the same location. It can lead to the creation of strong working groups or be used for solving urban concentration by overcoming spatial constraints.

\acknowledgments{%
This research was supported by the National Research Council of Science and Technology(NST) funded by the Ministry of Science and ICT (MSIT), Republic of Korea (No. CRC 21011, 50\%) and Institute of Information \& communications Technology Planning \& Evaluation (IITP) grant funded by the Korea government (MSIT) (No. RS-2024-00397663, Real-time XR Interface Technology Development for Environmental Adaptation, 50\%).
}

\bibliographystyle{abbrv-doi}

\bibliography{template}
\end{document}